\documentclass{article}
\usepackage{graphicx}
\usepackage{graphics}

\newcommand{\dfr}{d\raise0.3ex\hbox{\kern-0.5ex\char"013 }}

\renewcommand\baselinestretch 2
\begin{document}

\title{Chaotic motion in classical fluids with scale relativistic methods}
\author{Marie-No\"elle C\'el\'erier,\footnote{LUTH, Observatoire de Paris, CNRS,  Universit\'e Paris Diderot, 5 place Jules Janssen, 92195 Meudon cedex, France; e-mail: marie-noelle.celerier@obspm.fr}}
\maketitle

\abstract

In the framework of the scale relativity theory, the chaotic behavior in {\it time only} of a number of macroscopic systems corresponds to motion in a {\it space} with geodesics of fractal dimension 2 and leads to its representation by a Schr\"odinger-like equation acting {\it in the macroscopic domain}. The fluid interpretation of such a Schr\"odinger equation yields Euler and Navier-Stokes equations. We therefore choose to extend this formalism to study the properties of a system exhibiting a chaotic behavior {\it both in space and time} which amounts to consider them as issued from the geodesic features of a mathematical object exhibiting all the properties of a fractal `space-time'. Starting with the simplest Klein-Gordon-like form that can be given to the geodesic equation in this case, we obtain a motion equation for a `three fluid' velocity field and three continuity equations, together with parametric expressions for the three velocity components which allow us to derive relations between their non-vanishing curls. At the non relativistic limit and owing to the physical properties exhibited by this solution, we suggest that it could represent some kind of three-dimensional chaotic behavior in a classical fluid, tentatively turbulent if particular conditions are fulfilled. The appearance of a transition parameter ${\cal D}$ in the equations allows us to consider different ways of testing experimentally our proposal.




\section{Introduction}
\label{s.i}


The Madelung transformation of the Schr\"odinger equation has been long known as providing a tentative fluid-like interpretation of non relativistic quantum mechanics \cite{EM27,DB52}. When writing the complex wave function $\psi$ in polar form, $\psi=\sqrt{P} \rm{e}^{i\theta}$, then separating the real and imaginary parts of the Schr\"odinger equation, one obtains a Hamilton-Jacobi and a continuity equations where $P$ and $V= \nabla \theta$ can be interpreted as, respectively, the density and the velocity field of an irrotational flow in a compressible fluid. The gradient of the Hamilton-Jacobi equation is an Euler-type equation exhibiting a new term, the so-called `quantum' potential, which depends only on the density and has been, in the hydrodynamical picture, regarded as a pressure potential or a mechanical stress tensor \cite{TT52,TT53}.

In scale relativity, the effect on motion of the simplest scale laws, namely self-similar laws with fractal dimension 2, allows one to recover the various equations of quantum mechanics \cite{LN93,CN04,LN96b,CN06}. However, since the mathematical construction of the quantum equations thus obtained {\it does not depend on the value of the Planck constant $\hbar$} (see Sec.~\ref{s.qkg}), its results can be generalized to the classical realm provided some particular conditions are fulfilled by the system under consideration \cite{LN93,LN96b}. This has already been done for the Schr\"odinger equation with applications to the chaotic behavior in {\it time} of a number of macroscopic systems (planetary systems \cite{LN93,LN96a,NS97}, astrophysical structures \cite{NS98,DN03}, etc.) which have been successfully corroborated by observations.

Now, it must be remembered that it was shown that non-differentiability, which is at the root of the theory, implies a fractal space(-time) which in turn implies an infinite number of geodesics defined at any of its points, hence a fluid-like interpretation of such geodesic bundles \cite{LN93,CN04,CN03,NC07}. In the case where the fractality of space only is involved, the parameter along the geodesics is the time $t$ and the differentiable elements representing the resolutions which define the scale state of the system under consideration is ${\rm d} t$. A fluid interpretation of the macroscopic Schr\"odinger-type equation can therefore be developed which allows one to obtain {\it classical} fluid motion equations, i. e., Euler and Navier-Stokes equations, with a `quantum-like' potential \cite{LN09}.

To strengthen the appropriateness of the use of scale relativity methods to deal with chaotic motion, we stress that the quantum-type equations, from which the fluid motion equations are derived, are obtained in this framework when three conditions are fulfilled. They are consequences of the giving up of the differentiability assumption and read, in the simplest case when only the $dt \leftrightarrow - \, dt$ symmetry is broken \cite{LN93}:

1. The geodesic number is infinite as a consequence of the fractality of space. Thus, a fluid-like description obtains in which the velocity $v(t)$ is replaced by a velocity field $v[x(t),t]$.

2. The geodesics are fractal curves with fractal dimension 2 which is the dimension of a Markovian process that we take as a representation of the assumed chaotic behavior of our system.

3. Due to non-differentiability, the invariance under the differential element symmetry $dt \leftrightarrow - \, dt$ is broken, in the sense that the standard definition of a derivative where $dt$ is assumed to vanish does no more apply. One therefore replaces the coordinates $x(t)$ by fractal functions $x(t, dt)$ and the velocity becomes two-valued as
\begin{equation}
v_+[x(t,dt),t,dt] = \frac{x(t + dt,dt) - x(t,dt)}{dt},
\label{i1}
\end{equation}
\begin{equation}
v_-[x(t,dt),t,dt] = \frac{x(t,dt) - x(t - dt,dt)}{dt},
\label{i2}
\end{equation}
This yields two fractal velocity fields
\begin{equation}
V_+ = v_+[x(t),t] + w_+ [x(t,dt),t,dt],
\label{i3}
\end{equation}
\begin{equation}
V_- = v_-[x(t),t] + w_- [x(t,dt),t,dt],
\label{i4}
\end{equation}
where $v_+$, respectively $v_-$, is a `classical part', differentiable and independent of the resolution $dt$, and $w_+$, respectively $w_-$, is a `fractal part', explicitly dependent on $dt$ and divergent when $dt$ goes to zero. The `classical part' is shown to dominate over the `fractal part' above (or below) some transition scale characterized by a parameter ${\cal D}$ \cite{LN93,CN04}, hence two different behaviors of the system: non-chaotic beyond (below) this scale and chaotic below (beyond).

When one wants to go farther on, one must consider that the parameter along the geodesics is the proper time $s$ and thus, as a first step,  consider the breaking of the space-time differential element $ds$, which, when considered alone, gives the complex Klein-Gordon equation \cite{LN96b}. This amounts to analyze only the effect of non-differentiability on the total derivative $d/ds$. However, the velocity fields of the `geodesic' bundles are also functions of the coordinates, so that one is led to consider also the effect of the partial derivatives $\partial / \partial x^{\mu}$ in the decomposition $d/ds = \partial /\partial s + (dx^{\mu}/ds) \partial / \partial x^{\mu}$ which yields the new symmetry breaking $dx^{\mu} \leftrightarrow - \, dx^{\mu}$ \cite{CN04}. In this case, the corresponding algebra is that of the quaternions.

Note that, since we are now in a full space-time representation, the time $t$, which was a mere parameter on the geodesics in the previous case, becomes here a coordinate of a fractal four-dimensional space-time. Thus, while projecting the fractal geodesic curves on such a coordinate, we see that time becomes also a fractal function, as it is the case of the three spatial coordinates.

Therefore, we propose here to examine whether the chaotic behavior both in time {\it and} space of the macroscopic system we consider can be constructed as emerging from the breaking, at the level of an underlying mathematical object which have all the properties of a fractal `space-time', of {\it both} the differential element symmetries $ds \leftrightarrow - \, ds$ {\it and} $dx^{\mu} \leftrightarrow - \, dx^{\mu}$. The development of such a formalism, which yields a quaternionic Klein-Gordon equation, is followed by a transition to the non-relativistic limit, since our aim is to model non-relativistic flows. As a first step, we limit ourselves to the simplest motion equation which can be written in this framework and which corresponds to free (inertial) motion in the fractal `space-time'. We leave the study of more complicated behavior to future work. However, we are thus able to show that new and interesting properties emerge as well as a new way of characterizing the possible transition to some kind of `turbulence'.

A long-known property of the quaternionic formalism is that the Dirac equation for a free particle can be obtained as a mere square root of the Klein-Gordon operator \cite{CL29,AC37}. This property has been used to derive the Dirac equation as a geodesic equation in a non-differentiable therefore fractal space-time in the case when the velocity field of the geodesic fluid, and therefore the wave function, is represented by bi-quaternions (i. e., complex quaternions) \cite{CN04,CN03}. The bi-quaternionic wave function has been shown to exhibit all the properties of Dirac bi-spinors. Moreover, when one takes the non-relativistic limit of this bi-quaternionic equation, one recovers the Pauli equation and all the well known properties of the non-relativistic Pauli spinor \cite{CN06}. We therefore suspected, in a first stage of our study, that relativistic fluids whose motion equations could be obtained by such a process might have richer properties than those of the simple fluids obtained in the complex wave function case.

However, when trying to use this method in order to obtain some chaotic behavior for a classical fluid as the non-relativistic limit of a relativistic bi-quaternionic one, the vorticities of the velocity fields disappear and we are left with a set of gradients which are not well designed to represent a whirling fluid.

This comes from the way algebra doublings appear in this context (see Sec.~\ref{s.ckg}). Actually, to each symmetry breaking, $ds \leftrightarrow - \, ds$, $dx^{\mu} \leftrightarrow - \, dx^{\mu}$, then $x^{\mu} \leftrightarrow - \, x^{\mu}$ corresponds a doubling of the algebra, which allows one to use, first, the complex numbers, then the quaternions and finally the bi-quaternions, and their related algebras \cite{CN04}. However, the breaking of the symmetry $x^{\mu} \leftrightarrow - \, x^{\mu}$ is a specific property of the microscopic quantum domain which is not observed in classical physics. Therefore, we do not need to consider here the breaking of this symmetry and we are left with only two symmetry breakings and therefore with the quaternionic algebra.

A first interesting attempt was made in this context by Love and Boghosian \cite{LB04}. However, these authors consider a Madelung transformation of the quaternionic Schr\"odinger equation without giving any physical justification for this choice. Only a mathematical consideration is put forward to explain it. Note moreover that, in the scale relativity framework, the Schr\"odinger equation can be obtained from only one symmetry breaking, namely that of the differential time element $dt$. Therefore its corresponding algebra is that of the complex numbers, not of the quaternions.

The content of this paper is as follows. In Sec.~\ref{s.ckg}, we give a reminder of the method which allows one to recover the Klein-Gordon equation with a complex wave function and the standard Euler equation in the context of scale relativity. This is provided in order to give the reader a simple example of how the scale relativity methods can be used to obtain classical motion equations of fluid mechanics from quantum-like mechanical equations assumed to act in the macroscopic domain. In Sec.~\ref{s.qkg}, we jump to the quaternionic case and derive a set of motion equations for the three 4-vector components of a `relativistic' non-Abelian fluid velocity field, together with parametric expressions for these components which exhibit their main physical properties. In Sec.~\ref{s.nrl}, we take the non-relativistic limit of the solution obtained in Sec.~\ref{s.qkg} which gives motion equations for the fluid velocity field which appears under the form of a set of three vector fields. Three conservation equations and expressions for the non vanishing curls of these three vector fields are also obtained at this limit. Owing to the way we have constructed this solution from a scale relativistic treatment of chaos, we propose that, provided particular conditions are fulfilled, some kind of chaotic motion, possibly turbulent, might be represented by such a `three fluid' . A discussion and our conclusion are given in Sec.~\ref{s.concl}.


\section{Complex Klein-Gordon equation and Euler equation} 
\label{s.ckg}


Following Madelung, some authors have applied the fluid representation to the complex Klein-Gordon field of relativistic quantum mechanics \cite{TT52,TT53,LN05}. They also obtain an Euler-type equation for a relativistic fluid and a conservation equation for a relativistic current-type quantity $j^{\mu}$. However, one can derive directly the fluid motion equations from geodesic equations written with the scale relativity covariant derivative without any need to integrate them previously under the form of a Klein-Gordon equation \cite{LN05}. This is the method that we apply here.

The example of a scale relativity method implementation given in this Section is provided as a summarized reminder. Since a fully detailed development and discussion of all the aspects of the formalism is beyond the scope of this paper, we refer the interested reader to the original articles.

Scale relativity generalizes to scales the relativity principle only applied to position, orientation and motion in other standard relativity theories.

Here, space-time is assumed to be a continuum which is not only curved, as in general relativity, but can also be non-differentiable and therefore fractal \cite{LN93}. The infinite number of fractal geodesics at each point of such a fractal space-time allows a fluid representation of the theory (see \cite{NC07} for a discussion). The effect on motion of the simplest scale laws, namely self-similar laws with fractal dimension 2, allows one to recover the various equations of quantum mechanics \cite{CN04,LN96b,CN06}. They are derived as geodesic equations each corresponding to a given level of discrete symmetry breaking, namely of the reflection invariances of differential elements, which are implicit symmetries of classical mechanics.

The Schr\"odinger equation is obtained as the integral of a geodesic equation acting in a fractal space and issued from the breaking of the reflection invariance of the time differential element $dt \leftrightarrow - \, dt$ \cite{LN93}. The Klein-Gordon equation proceeds from a geodesic equation acting in the fractal space-time and from the breaking of the relativistic line element symmetry $ds\leftrightarrow - \, ds$ \cite{LN96b,LN94b}. In both cases the wave function is complex as a consequence of the two-valuedness issued from these symmetry breakings.

Then, a new two-valuedness is introduced, in the motion-relativistic case, from the breaking of the discrete symmetry of the space-time coordinate differential elements $dx^{\mu} \leftrightarrow - \, dx^{\mu}$. It is naturally described in terms of quaternions (spinors). Adding parity violation, i. e., the breaking of the discrete symmetry $x^{\mu} \leftrightarrow - \, x^{\mu}$, which is a property of standard quantum theory, one is led to introduce bi-quaternions (bi-spinors). The Dirac equation is then derived as the bi-quaternionic square root of a bi-quaternionic Klein-Gordon equation issued from a geodesic equation acting in the fractal space-time \cite{CN04,CN03}. The Pauli equation is recovered as the quaternionic non-relativistic limit of this Dirac equation \cite{CN06}.

The fluid interpretation of the Schr\"odinger equation in the framework of scale relativity was developed in \cite{LN09,NC07,LN05}. A generalization to the relativistic complex Klein-Gordon case was given in \cite{LN05}. We recall briefly below the main steps of the reasoning since it will be adapted to the quaternionic case in Sec.~\ref{s.qkg}. Instead of performing a Madelung-type transformation of the Klein-Gordon equation, the idea is to take advantage of the formulation of this equation as a geodesic equation in the fractal space-time to obtain the fluid motion equation directly from the covariant equations of the dynamics.

We have seen in Sec.~\ref{s.i} that, in the scale relativity approach, the velocity field on the geodesic bundle of the non-differentiable space(-time) is decomposed in terms of a differentiable part, ${\cal V}$, and of a fractal, divergent, non-differentiable part, ${\cal W}$ (of zero mean). At this stage of the construction, both are complex due to the fundamental two-valuedness issued from the breaking of the symmetry $ds\leftrightarrow - \, ds$. A complex covariant total derivative can be built from the differentiable part of the velocity field ${\cal V}$ which, when applying to the relativistic case, is \cite{CN04,LN94b}
\begin{equation}
\frac{\dfr}{ds} = \left({\cal V}^{\mu} + i\frac{\lambda}{2}\partial^{\mu}\right) \partial_{\mu},
\label{ckg5}
\end{equation}
where the constant $\lambda$ gives the amplitude of the fractal fluctuations $d\xi$, as $\lambda = \langle d\xi^2 \rangle /ds$ (in the quantum mechanics interpretation, $\lambda$ is the Compton length of the `particle', $\lambda = \hbar/mc$). Now one assumes that the system can be characterized by an action ${\cal S}$ which is complex such as the four-velocity. Following the same reasoning as in classical mechanics, one writes
\begin{equation}
mc \; {\cal V}_{\nu}= -\partial_{\nu}{\cal S}.
\label{ckg5b}
\end{equation}
The stationary action principle applied on this action yields a free equation of motion which takes the following form of a geodesic equation
\begin{equation}
\frac{\dfr {\cal V}_{\nu}}{ds} = 0.
\label{ckg6}
\end{equation}
Then, a `wave-function' is introduced, which is nothing but another expression of the complex action, and which reads
\begin{equation}
\psi    =   {\rm e}  ^{i{\cal S} /S_0 },
\label{ckg6b}
\end{equation}
where $S_0$ is a constant, needed to ensure the correct dimensionality of the equation and which can be written $S_0 = m c \lambda$. Therefore, the complex velocity field becomes
\begin{equation}
{\cal V} _{\nu } = {i \lambda} \partial_{\nu }\ln\psi.
\label{ckg6c}
\end{equation}
Introducing the `wave-function' in the motion equation Eq.~(\ref{ckg6}), one obtains
\begin{equation}
\left(\partial^{\mu} \ln \psi + \frac{1}{2} \partial^{\mu} \right) \partial_{\mu} \partial_{\nu} \ln \psi = 0,
\label{ckg6d}
\end{equation}
which, after some algebra, takes the form
\begin{equation}
\partial ^{\nu } \left(  \frac{\partial ^{\mu }\partial _{\mu }\psi}{\psi} \right)     =  0.
\label{ckg6e}
\end{equation}
It can be therefore integrated as a Klein-Gordon equation, which reads
\begin{equation}
\partial ^{\mu } \partial _{\mu}\psi + \frac{\psi}{\lambda^2} = 0.
\label{ckg6f}
\end{equation}
Such a construction can be generalized to the case when the full velocity field ${\cal V} + {\cal W}$ is taken into account in the motion equation and in the covariant derivative. However, this leads to the same results as regards the final quantum mechanical-type equations \cite{LN99}.

Now, we come back to Eq.~(\ref{ckg6}). The complex velocity is decomposed into its real and imaginary parts, ${\cal V}_{\nu} = V_{\nu} -iU_{\nu}$, so that the geodesic equation becomes
\begin{equation}
\left[V^{\mu} - i\left(U^{\mu} - \frac{\lambda}{2}\partial^{\mu}\right)\right] \partial_{\mu}(V_{\nu} - iU_{\nu}) = 0.
\label{ckg7}
\end{equation}
The real part of this equation takes the form of a relativistic Euler equation which reads
\begin{equation}
V^{\mu}\partial_{\mu}V_{\nu} = \frac{d V_{\nu}}{ds} = \left(U^{\mu} - \frac{\lambda}{2}\partial^{\mu}\right) \partial_{\mu}U_{\nu}.
\label{ckg8}
\end{equation}
Equation (\ref{ckg6c}) gives the imaginary part of the complex velocity field ${\cal V}_{\nu}$, in terms of the wave function modulus, $\sqrt{P}$, under the form
\begin{equation}
U_{\nu} = - \lambda \partial_{\nu} \ln \sqrt{P}.
\label{ckg10}
\end{equation}
Inserting this expression into the right-hand side of Eq.~(\ref{ckg8}), one obtains, after some calculations, the following equation of dynamics \cite{LN05}
\begin{equation}
\frac{d V_{\nu}}{ds} = \frac{F_{\nu}}{m} = \frac{1}{2} \lambda^2 \partial_{\nu} \left(\frac{\partial^{\mu} \partial_{\mu}\sqrt{P}}{\sqrt{P}}\right).
\label{ckg11}
\end{equation}
Another way of writing Eq.~(\ref{ckg11}) is
\begin{equation}
V^{\mu} \partial_{\mu} V_{\nu} = \frac{1}{2} \lambda^2 \partial_{\nu} \left(\frac{\partial^{\mu} \partial_{\mu}\sqrt{P}}{\sqrt{P}}\right),
\label{ckg12c}
\end{equation}

Now, the complex velocity, ${\cal V}_{\nu}=V_{\nu}-iU_{\nu}$, being linked to the `wave-function' by Eq.~(\ref{ckg6c}), its real part writes, in term of the phase,
\begin{equation}
V_{\nu} = - \frac{\lambda}{S_0}\partial_{\nu} S,
\label{ckg13}
\end{equation}
where $S$ is the real part of the complex action ${\cal S}$.

At the non-relativistic limit ($c \rightarrow \infty$), the Dalembertian $\partial^{\mu} \partial_{\mu} = \partial^2/c^2\partial t^2 - \Delta$ reduces to $-\Delta$, the four-derivative $\partial_{\nu}$ becomes $-\nabla$ and the fractal parameter $\lambda$ becomes $2{\cal D} = c \lambda$ \footnote{Actually, the relativistic definition of $\lambda$ is: $\langle d\xi_{\pm}^{\mu} d\xi_{\pm}^{\nu} \rangle = \mp \lambda \eta^{\mu \nu} ds$ and the non-relativistic definition of ${\cal D}$ is $\langle d\xi_{\pm i} d\xi_{\pm j} \rangle = \pm 2 {\cal D} \delta_{ij} dt$, where $\eta^{\mu \nu}$ is the Minkowski metric tensor and $\delta_{ij}$ is the Kronecker symbol. Since $\eta^{ij} = - \delta_{ij}$ and in proper time $ds = c dt$, at the non-relativistic limit $\lambda$ becomes $c \lambda = 2{\cal D}$. This can be readily verified by a dimensional analysis of the equations.}. Since in the quantum domain $\lambda = \hbar / m c$, we recover the non-relativistic standard `quantum potential' ${\cal Q} = - \hbar^2/2m \, \Delta \sqrt{P}/ \sqrt{P}$, which can be interpreted as a byproduct of the underlying geometry of the fractal space \cite{NC07,LN09,LN05}.
At this limit Eq.~(\ref{ckg12c}) becomes
\begin{equation}
\left( \frac{\partial}{\partial t} + V . \nabla \right) V = - 2 {\cal D}^2 \nabla \left( \frac{\Delta \sqrt{P}}{\sqrt{P}} \right),
\label{ckg12d}
\end{equation}
and Eq.~(\ref{ckg13}) becomes
\begin{equation}
V =\frac{\nabla S}{m},
\label{ckg14}
\end{equation}
We recognize in Eq.~(\ref{ckg12d}) the standard Euler equation with a `quantum potential' obtained when performing a Madelung transformation of the Schr\"odinger equation with no exterior potential and in Eq.~(\ref{ckg14}) the standard definition of the Madelung fluid velocity as a gradient of the phase of the wave-function. Note, however, that Eq.~(\ref{ckg14}) is here derived, since $V_{\nu}$ has been defined from the very beginning as the real part of the geodesic mean velocity field.

Finally, the continuity equation is recovered as the non-relativistic limit of the imaginary part of the Klein-Gordon equation. From the definition of the wave function given by Eq.~(\ref{ckg6b}), we obtain
\begin{equation}
\psi = \sqrt{P} {\rm e}^{iS/S_0},
\label{ckg15}
\end{equation}
where $S$ is the real part of the complex action, ${\cal S}$, and $\sqrt{P}$ comes from its pure imaginary part. Now, we calculate $\partial^{\mu}\partial_{\mu}\psi$ which we substitute in the Klein-Gordon equation (\ref{ckg6f}). Extracting the pure imaginary part and using Eq.~(\ref{ckg13}), we obtain the continuity equation under the form
\begin{equation}
\partial^{\mu}(PV_{\mu}) = 0,
\label{ckg16}
\end{equation}
which, at the non-relativistic limit, becomes
\begin{equation}
\frac{\partial P}{\partial t} + \nabla (PV) = 0.
\label{ckg17}
\end{equation}
This implies that the square of the wave-function modulus, $P$, can be regarded as linked to a genuine density, $\rho$, by $mP = \rho$, in the fluid interpretation of the space-time geodesics. Actually, in the scale relativity space-time approach, the probability density characterizes the potential geodesic bundle from the very beginning of the description. In this interpretation, the geodesic fluid is expected to be more concentrated somewhere and less concentrated elsewhere. It fills some regions and is nearly vanishing in others, as does a fluid. This behavior can be described by a presence probability density of the paths of the fluid elements and reduces to fluid motion equations in the classical domain where the classical part of the velocity field happens to dominate the fractal fluctuations \cite{NC07,LN09}.


\section{Quaternionic Klein-Gordon-type equation} 
\label{s.qkg}


In the following, we will consider the behavior of a system which is chaotic both in space {\it and} time. Such a property has often been ascribed to turbulence \cite{MR97}. However, since a proper characterization of turbulence is much involved, we want to be cautious about the interpretation of our results which will be discussed in Sec.~\ref{s.concl}.

First, we want to stress that, even if the mathematical construction of the following is the same as the mathematical construction of the bi-quaternionic Klein-Gordon equation from which the Dirac equation is derived, its physical interpretation is different. Actually, we do not intend to claim that the macroscopic fluid whose features might emerge from our analysis will possess `quantum' properties, neither would it be relativistic. This should be forbidden by the non-physical nature of the quaternionic Klein-Gordon-type equation which is unable to yield a Dirac equation for bi-spinors. Now, it is long known from experiment that the actual nature of the electron-positron couple is of the bi-spinor type. Therefore we consider the following quaternionic construction as typical of a representation of chaotic motion in space and time which is merely relevant {\it at the non-relativistic limit} and {\it in the macroscopic domain}.

We want to stress once more that, in the scale relativity framework, quantum-like equations have been proposed to act in the macroscopic domain, providing therefore the constant $\lambda$ with a new status independent of the value of $\hbar$. Actually, it is easy to verify that the whole mathematical structure of the theory does not rely on $\hbar$ and that it is preserved for any value of $\lambda$ which is introduced here as a transition length scale \cite{CN03}. This was first used for applications issued from the Schr\"odinger equation \cite{LN93,LN96b}, but it also obtains in the construction completed here. We therefore propose that Eqs.~(\ref{nrl1}) to (\ref{nrl7}) describe the behavior of a macroscopic non-relativistic fluid whose properties we examine below.

Using the scale relativity method for deriving motion equations in a fractal space-time as recalled in Sec.~\ref{s.ckg}, we introduce a quaternionic `wave function', $\psi$, which is defined as another expression for the quaternionic action and reads
\begin{equation}
\psi^{-1}\partial_{\mu}\psi \equiv \frac{e_1}{S_0} \partial_{\mu}S = - \frac{e_1 mc}{S_0} {\cal V}_{\mu},
\label{qkg1}
\end{equation}
where ${\cal V}_{\mu}$ is the differentiable part of the quaternionic velocity field and $S_0 / mc$ is a constant we rename $\lambda$ since it plays the same role of a transition scale as in the complex case. Here, $e_1$ is any of the three basis elements of the quaternionic algebra, $\{e_1,e_2,e_3\}$, which satisfy $e_1^2 = e_2^2 = e_3^2 = e_1e_2e_3 = -1$. It is, for this algebra, an analog of the imaginary $i$ for the complex number algebra. This is the reason why we use it to define the quaternionic `wave function' from the quaternionic action.

We want also to stress that the `wave function', $\psi$, is here a mere mathematical tool used to express the action from which the velocity is defined. Since we do not consider this reasoning as valid to describe quantum mechanical processes we might have given to this tool any other different name.

Since $\psi^{-1}\psi=1$, which implies $\psi^{-1} \partial_{\mu}\psi = -\partial_{\mu}\psi^{-1} . \psi$, we can write
\begin{equation}
{\cal V}_{\mu} = - \, e_1 \, \lambda \, \partial_{\mu}\psi^{-1} . \psi.
\label{qkg2}
\end{equation}

In the following, we will first compute the quaternionic components of the velocity field ${\cal V}_{\mu}$ under a parametric form to exhibit some of their physical properties, then we will employ the method of \cite{LN05} to derive directly the fluid motion equation from a free geodesic equation and finally we  will extract the quaternionic imaginary parts of the Klein-Gordon-type equation to obtain conservation equations for the currents.

For the sake of calculation simplification, we choose to write the quaternionic wave function as
\begin{equation}
\psi = \sqrt{P} g^{-1},
\label{qkg3}
\end{equation}
where $g$ is any unimodular quaternion, and therefore such is $g^{-1}$. Inserting this expression in Eq.~(\ref{qkg2}), we obtain
\begin{equation}
{\cal V}_{\mu} =  e_1 \left(\frac{\lambda}{2} \, \partial_{\mu} \ln P - \lambda \, \partial_{\mu}g.g^{-1} \right).
\label{qkg4}
\end{equation}

Since $g$ is unimodular, we can write it under the form
\begin{equation}
g = {\rm exp} (\alpha e_1 + \beta e_2 + \gamma e_3),
\label{qkg5}
\end{equation}
where $\alpha, \beta$ and $\gamma$ are real fields. From this definition of $g$ we obtain an expression for $\partial_{\mu}g.g^{-1}$ which we insert into Eq.~(\ref{qkg4}). After some calculations using the quaternionic algebra, we obtain a decomposition of ${\cal V}_{\mu}$ into its quaternionic components which are functions of the parameters $\alpha, \beta$ and $\gamma$.

If we adopt the following notation for these components
\begin{equation}
{\cal V}_{\mu} = v_{0 \mu} + v_{1 \mu} e_1 + v_{2 \mu} e_2 + v_{3 \mu} e_3,
\label{qkg6}
\end{equation}
their parametric expressions are
\begin{equation}
v_{0 \mu} = \lambda (\partial_{\mu} \alpha + \sin 2 \beta \, \partial_{\mu} \gamma),
\label{qkg7}
\end{equation}
\begin{equation}
v_{1 \mu} = \frac{\lambda}{2} \partial_{\mu} \ln P,
\label{qkg8}
\end{equation}
\begin{equation}
v_{2 \mu} = \lambda (\sin 2 \alpha \, \partial_{\mu} \beta + \cos 2 \alpha \cos 2 \beta \, \partial_{\mu} \gamma),
\label{qkg9}
\end{equation}
\begin{equation}
v_{3 \mu} = - \lambda (\cos 2 \alpha \, \partial_{\mu} \beta - \sin 2 \alpha \cos 2 \beta \, \partial_{\mu} \gamma).
\label{qkg10}
\end{equation}
It is easy, using this parametric representation, to check that the curls of the three fields $v_0, v_2$ and $v_3$ are non zero and verify
\begin{equation}
\partial_{\mu}v_{0 \nu} - \partial_{\nu}v_{0 \mu} = \frac{2}{\lambda}(v_{2 \mu} v_{3 \nu} - v_{2 \nu} v_{3 \mu}),
\label{qkg11}
\end{equation}
\begin{equation}
\partial_{\mu}v_{2 \nu} - \partial_{\nu}v_{2 \mu} = \frac{2}{\lambda}(v_{3 \mu} v_{0 \nu} - v_{3 \nu} v_{0 \mu}),
\label{qkg12}
\end{equation}
\begin{equation}
\partial_{\mu}v_{3 \nu} - \partial_{\nu}v_{3 \mu} = \frac{2}{\lambda}(v_{0 \mu} v_{2 \nu} - v_{0 \nu} v_{2 \mu}).
\label{qkg13}
\end{equation}
As regards $v_1$, which appears as a gradient of the logarithm of $P$, it will give rise below to the `quantum' potential term in the motion equation.

The quaternionic covariant total derivative applying here is given by Eq.~(\ref{ckg5}) where the complex imaginary $i$ is replaced by the quaternionic imaginary $e_1$ as above (see \cite{CN04,CN03} for an equivalent detailed derivation in the bi-quaternionic case). This gives
\begin{equation}
\frac{\dfr}{ds} = \left({\cal V}^{\mu} + e_1 \frac{\lambda}{2}\partial^{\mu}\right) \partial_{\mu}.
\label{qkg14}
\end{equation}
Then, we insert this expression together with the decomposition of the velocity field ${\cal V}$, given by Eq.~(\ref{qkg6}) where $v_1$ has been replaced by its expression given by Eq.~(\ref{qkg8}), into the free (inertial) motion equation (\ref{ckg6}). After some rearrangements, we obtain
\begin{equation}
v_0^{\mu} \partial_{\mu} v_{0 \nu} - v_2^{\mu} \partial_{\mu} v_{2 \nu} - v_3^{\mu} \partial_{\mu} v_{3 \nu} = \frac{\lambda^2}{2} \partial_{\nu} \left(\frac{\partial^{\mu} \partial_{\mu} \sqrt{P}}{\sqrt{P}} \right).
\label{qkg15}
\end{equation}
Now, for the sake of mathematical simplification, and as it has been done in Sec.~\ref{s.ckg}, we derive the three other equations corresponding to the pure imaginary terms of the quaternionic motion equation from the Klein-Gordon-type equation, i. e., from the equation obtained when integrating Eq.~(\ref{ckg6}). This yields already integrated equations under the form of continuity equations which are more easy to interpret.

The quaternionic Klein-Gordon-type equation can be obtained with exactly the same reasoning employed to obtain it in the bi-quaternionic case \cite{CN04,CN03}. It reads
\begin{equation}
\partial^{\mu} \partial_{\mu} \psi + \frac{1}{\lambda^2} \psi = 0,
\label{qkg16}
\end{equation}
where $\psi$ is the `wave function' expressed by Eq.~(\ref{qkg3}).

We first expand $\partial^{\mu} \partial_{\mu} \psi$ as
\begin{equation}
\partial^{\mu} \partial_{\mu} \psi = \partial^{\mu} \partial_{\mu} \sqrt{P} . g^{-1} + 2 \, \partial^{\mu} \sqrt{P} . \partial_{\mu} g^{-1} + \sqrt{P} \, \partial^{\mu} \partial_{\mu} g^{-1}.
\label{qkg16b}
\end{equation}
Then, remarking that $\partial_{\mu} g^{-1} = - g^{-1}(\partial_{\mu} g . g^{-1})$ and using Eqs.~(\ref{qkg4}), (\ref{qkg6}) and (\ref{qkg8}), we obtain, after some algebra, an expression for $\partial^{\mu} \partial_{\mu} \psi$ which we insert into Eq.~(\ref{qkg16}) where $\psi$ has been replaced by its expression given by Eq.~(\ref{qkg3}). Some final calculations give the Klein-Gordon equation under the form
\begin{eqnarray}
&&\lambda^2 \partial^{\mu} \partial_{\mu} \sqrt{P} - \sqrt{P} (v_0^2 + v_2^2 + v_3^2) + \sqrt{P} - 2 \lambda \, \partial^{\mu} \sqrt{P} (v_{0 \mu} e_1   \\ \nonumber
&&- v_{3 \mu} e_2 +  v_{2 \mu} e_3)
- \lambda  \sqrt{P} (\partial^{\mu} v_{0 \mu} e_1 - \partial^{\mu} v_{3 \mu} e_2 + \partial^{\mu} v_{2 \mu} e_3) = 0.
\label{qkg17}
\end{eqnarray}
We extract each component of the quaternionic imaginary terms in the lhs of Eq.~(70) and equate them to zero. After some arrangements, we obtain three continuity equations which read
\begin{equation}
\partial^{\mu} (P v_{0 \mu}) = 0,
\label{qkg18}
\end{equation}
\begin{equation}
\partial^{\mu} (P v_{2 \mu}) = 0,
\label{qkg19}
\end{equation}
\begin{equation}
\partial^{\mu} (P v_{3 \mu}) = 0.
\label{qkg20}
\end{equation}


\section{Non-relativistic limit} \label{s.nrl}


At the non-relativistic limit, $c \rightarrow \infty$, the time component of the quaternionic velocity is ${\cal V}^0 = c$ \cite{CN06} and the fractal parameter $\lambda$ becomes $2 {\cal D}$ (see the footnote in Sec.~\ref{s.ckg}). Moreover, since our purpose is to use the mathematical construction described here to apply it to the derivation of motion equations for fluids in the macroscopic domain, this fractal parameter is no more linked to $\hbar$ but it is supposed to take macroscopic values. Therefore, Eq.~(\ref{qkg15}), where we define the fluid density $\rho = mP$, becomes
\begin{equation}
\left( \frac{\partial}{\partial t} + v_0 . \nabla \right) v_0 - \left( \frac{\partial}{\partial t} + v_2 . \nabla \right) v_2 - \left( \frac{\partial}{\partial t} + v_3 . \nabla \right) v_3 = - 2{\cal D}^2 \nabla \left( \frac{\Delta \sqrt{\rho}}{\sqrt{\rho}} \right),
\label{nrl1}
\end{equation}
which can be regarded as a motion equation for a `three fluid' with velocity field $V = \{v_0, v_2, v_3\}$, where the $v_i$'s are vector fields in 3-space, and with `quantum' potential
\begin{equation}
{\cal Q} = 2 m {\cal D}^2 \frac{\Delta \sqrt{\rho}}{\sqrt{\rho}}.
\label{nrl1b}
\end{equation}
Note that the sign of this potential is arbitrary since we could have chosen a reversed sign for the lhs of Eq.~(\ref{nrl1}). We remark that the quantity ${\cal D}$ has the dimension of a diffusion coefficient. Moreover, it is this parameter which determines the transition from chaotic to non-chaotic behavior.

Now, we take the non-relativistic limit of Eqs.~(\ref{qkg11}) to (\ref{qkg13}) and we obtain the following relations between the curls of each of the velocity field component and the cross product of the other two:
\begin{equation}
\nabla \times v_0 = \frac{1}{\cal D} \, v_2 \times v_3,
\label{nrl2}
\end{equation}
\begin{equation}
\nabla \times v_2 = \frac{1}{\cal D} \, v_3 \times v_0,
\label{nrl3}
\end{equation}
\begin{equation}
\nabla \times v_3 = \frac{1}{\cal D} \, v_0 \times v_2.
\label{nrl4}
\end{equation}
We recover here relations analogous to those found in \cite{LB04}, with an additional intervention of the coefficient ${\cal D}$ which naturally arises in our formalism and gives the equations the correct dimension. They show that, in the most general configuration, the three fields have non zero curls which are each perpendicular to the other two fields and that these curl amplitudes are inversely proportional to ${\cal D}$ and therefore to the strength of the potential.

Now, we remark that the cases when $v_i = v_j \times v_k$, for $i,j,k = 0,2,3$ and circular permutations of the indexes are particular solutions exhibiting the properties of Beltrami-type flows. In these cases, Eqs.~(\ref{nrl2}) to (\ref{nrl4}) give $\nabla \times v_i = {\cal D}^{-1} v_i$ for $i = 0,2,3$ and the ABC flow relations read $A_i= B_j C_k - B_k C_j$, $B_i = C_jA_k - C_k A_j$, $C_i = A_j B_k - A_k B_j$ and six other equations with circular permutations of the indexes. We recognize the characteristic properties of Beltrami flows which are known to be chaotic \cite{AK99}. Since such flows are particular cases of ours, this confirms that our equations describe also some chaotic motion. However, Beltrami flows exhibit a vanishing velocity divergence which is not the case of our general solution. We can therefore conclude that our equations are able to describe richer behaviors.

Finally, we take the non-relativistic limit of Eqs.~(\ref{qkg18}) to (\ref{qkg20}) and obtain
\begin{equation}
\frac{\partial \rho}{\partial t} + \nabla (\rho v_0) = 0,
\label{nrl5}
\end{equation}
\begin{equation}
\frac{\partial \rho}{\partial t} + \nabla (\rho v_2) = 0,
\label{nrl6}
\end{equation}
\begin{equation}
\frac{\partial \rho}{\partial t} + \nabla (\rho v_3) = 0,
\label{nrl7}
\end{equation}
which are continuity equations for the three fluid `currents'. Under this form, these equations apply to compressible fluids, with density $\rho(t,x,y,z)$. For incompressible fluids, $\rho= const.$, they become
\begin{equation}
\nabla v_0 = 0,
\label{nrl5b}
\end{equation}
\begin{equation}
\nabla v_2 = 0,
\label{nrl6b}
\end{equation}
\begin{equation}
\nabla v_3 = 0,
\label{nrl7b}
\end{equation}
while the `quantum' potential in Eq.~(\ref{nrl1}) vanishes and Eqs.~(\ref{nrl2}) to (\ref{nrl4}) remain unchanged.

Having found the general equations issued from our construction, we want now to discuss briefly some particular solutions which might be of interest in peculiar cases. We first equate the expressions of $\partial \rho/\partial t$ extracted from Eqs.~(\ref{nrl5}) to (\ref{nrl7}) and obtain
\begin{equation}
\nabla (\rho v_0) = \nabla (\rho v_2) = \nabla (\rho v_3).
\label{nrl8}
\end{equation}
The first above equation gives
\begin{equation}
\frac{\nabla (v_0 - v_2)}{v_0 - v_2} + \frac{\nabla \rho}{\rho} = 0.
\label{nrl9}
\end{equation}
A particular solution of this equation can be obtained after a separation of the variables and reads
\begin{equation}
v_{0i} - v_{2i} = \frac{a_i}{\rho},
\label{nrl10}
\end{equation}
where $i=x,y,z$ and the $a_i$s are integration constants. Analogous solutions can be found for the couples $\{v_2,v_3\}$ and $\{v_3,v_0\}$, which we write
\begin{equation}
v_{2i} - v_{3i} = \frac{b_i}{\rho},
\label{nrl11}
\end{equation}
\begin{equation}
v_{3i} - v_{0i} = \frac{c_i}{\rho}.
\label{nrl12}
\end{equation}
Eliminating $\rho$ we obtain
\begin{equation}
\frac{v_{0i} - v_{2i}}{a_i} = \frac{v_{2i} - v_{3i}}{b_i} = \frac{v_{3i} - v_{0i}}{c_i}
\label{nrl13}
\end{equation}
Since the $a_i$'s, $b_i$'s and $c_i$'s are constants (and not fields), we remark that the ratios of the differences between the velocity fields are constant in every direction, which is a further constraint pertaining only to this particular solution.

Let us stress, for completeness, that in the more particular case when $a_i=b_i=c_i$ for all $i$, we can have two cases. Either $v_0 = v_2$ (equivalently $v_0 = v_3$) and therefore Eq.~(\ref{nrl1}) becomes
\begin{equation}
\left(\frac{\partial}{\partial t} + v . \nabla \right) v = 2 {\cal D}^2 \nabla \left( \frac{\Delta \sqrt{\rho}}{\sqrt{\rho}} \right),
\label{nrl14}
\end{equation}
where we have renamed $v$ the field $v_3$ (equivalently $v_2$). In this case the curl of $v$ given by Eq.~(\ref{nrl4}) (equivalently Eq.~(\ref{nrl3})) vanishes and we recover a motion equation for an irrotational fluid with velocity $v$, subjected to the `quantum potential' ${\cal Q}$ with a minus sign. This is analogous to Eq.~(\ref{ckg12d}), i. e., the standard Madelung transformed Euler-type equation with no exterior potential, but with a reverse sign for the `quantum potential'.

The second case is obtained when $v_2 = v_3$. Then, Eq.~(\ref{nrl1}) becomes
\begin{equation}
\left(\frac{\partial}{\partial t} + v_0 . \nabla \right) v_0 -2\left(\frac{\partial}{\partial t} + v_2 . \nabla \right) v_2 = - 2 {\cal D}^2 \nabla \left( \frac{\Delta \sqrt{\rho}}{\sqrt{\rho}} \right).
\label{nrl15}
\end{equation}
Since $v_2 = v_3$, the curl of $v_0$ given by Eq.~(\ref{nrl2}) vanishes. We are left with a two component velocity field $V = \{v_0, v_2\}$ whose first component is a gradient and the second one is rotational. Note the appearance in Eq.~(\ref{nrl15}) of a factor 2 in front of the second term in the lhs which cannot be inserted in a redefinition of the velocity field $v_2$.

Now we come back to the general equations obtained in this section. Remember that the chaotic to non-chaotic transition is characterized by the macroscopic parameter ${\cal D}$. It was proposed, when applying the scale relativity tools to macroscopic systems, that ${\cal D}$ is a relative quantity specific of each system and that a generalized Heisenberg relation could give an order of its magnitude as $\delta x \, \delta v \simeq 2 {\cal D}$, where $\delta v$ is a velocity dispersion and $\delta x$ a position dispersion \cite{DD04}. Now, it is well known, from experiment, that a turbulent behavior occurs in a fluid with kinematic viscosity coefficient $\nu$ when its Reynolds number $Re \equiv V L / \nu$, where $V$ and $L$ are characteristic velocity and length scales of the flow field, becomes larger than some critical value $Re_c = V_t L_t/\nu$. It would be therefore interesting to check experimentally if $Re_c$ could be related to ${\cal D}$ by $2 {\cal D} = V_t L_t = Re_c \nu$. If this condition appeared to be fulfilled, it might be hints that some kind of turbulent flow in a fluid could be represented by a three component velocity field $V= \{v_0,v_2,v_3\}$ obeying a motion equation given by Eq.~(\ref{nrl1}) and three continuity equations, (Eqs.~(\ref{nrl5}) to (\ref{nrl7})). The curls of the three vector fields, $v_0,v_2,v_3$, would be non-vanishing and each of them would be parallel to the cross product of the other two fields. However, more investigations would be needed to confirm the turbulent behavior of a fluid with such characteristics.

It is here interesting to remark that the spontaneous appearance of vorticity at the macroscopic level for a chaotic behavior in space and time can be viewed as corresponding to the spontaneous appearance of spin in quantum mechanics from the fractal feature of space-time \cite{CN06}. Note however that both do not emerge at the same level of the construction since the spinor corresponds to the wave function $\psi$ and chaotic-vorticity to the derivative level expressed by Eq.~(\ref{qkg2}).


\section{Discussion and conclusion} \label{s.concl}


We have chosen to study the chaotic behavior of fluid motion with the methods of scale relativity in a fractal space-time corresponding to a `space-time chaos', since they have already proved to provide a good framework for the characterization of simpler fluid behaviors.

Using the same line of reasoning as that developed in \cite{LN05}, we have obtained a quaternionic Klein-Gordon-like equation, which, at the non-relativistic limit, yields a representation of a classical fluid behavior more complex than those previously obtained from a mere Schr\"odinger-type equation \cite{LN09}.

For the simplest case considered here, this solution is composed of a motion equation for a three component velocity field submitted to a `quantum' potential, together with continuity equations for the three fluid `currents'. The curls of each velocity component are non-vanishing and parallel to the cross product of the other two fields. We have also shown that particular solutions of our equations possess the physical properties of Beltrami flows which are known as exhibiting chaotic behaviors. Since our general solution is more complex than such simple flows, we propose it could be able to represent some kind of richer chaotic behavior, possibly turbulent.

Now, in our model, the chaotic to non-chaotic transition is characterized by a macroscopic parameter ${\cal D}$, which has the dimension of a diffusion coefficient. If some kind of turbulent behavior happened to be reproduced by our solution, it might be linked naturally, in our framework, to some `critical Reynolds number', $Re_c$, by $2 {\cal D} = V_t L_t = Re_c \nu$, where $\nu$ is the fluid kinematic viscosity coefficient.

We have thus all the tools to examine the way the transition from laminar to chaotic (possibly turbulent) behavior could occur in our representation. Actually, the extended chaotic behavior we propose can be viewed as the fundamental state emerging from the basic statement of our construction, i. e., the full relaxation of the differentiability assumption that leads to the breakings of the symmetries $ds \leftrightarrow - \, ds$ {\it and} $dx^{\mu} \leftrightarrow - \, dx^{\mu}$. The chaotic time behavior is the degenerate state when only the symmetry $dt \leftrightarrow - \, dt$ is broken. It yields, in the simplest inertial case considered here, a linear macroscopic Schr\"odinger equation which gives an Euler equation for an irrotational fluid that is therefore unable to exhibit a chaotic behavior.

However, there is no a priori reason for the parameter ${\cal D}_l$ yielding the Schr\"odinger equation from which a laminar behavior occurs to have the same value as the parameter ${\cal D}_t$ giving a Klein-Gordon equation from which a chaotic behavior can emerge, even for the same fluid. We might suspect that the length scale related to the former is of the order of the mean free path of the fluid molecules, while we know from experiment that the transition to chaos occurs at much larger scales. Therefore ${\cal D}_t$ should be considered as defining the scales characterizing the transition from non-chaotic to chaotic behavior, and possibly, the critical Reynolds number.

We have also shown that ${\cal D}_t$ appears both in the `quantum' potential expression and in the curls of the velocity field components. This allows us to propose three means of testing our results.

Since the strength of the `quantum potential' is proportional to ${\cal D}_t^2$, one might measure ${\cal D}_t$ in the chaotic phase of a given fluid subjected to such a potential. Then one could verify if the transition from a chaotic state to a laminar behavior occurs around some `critical Reynolds number' corresponding to ${\cal D}_t$.

A second test proposal is linked to our result that the amplitude of the velocity component curls are inversely proportional to ${\cal D}$. It is therefore a prediction of our representation that the greater the `quantum' or dissipative potential applied to the fluid, the less whirling and therefore the less chaotic it might be. This seems consistent with the observation that, e. g., spontaneous turbulence disappears when viscous dissipation erases it. However, it would deserve a more exact experimental confirmation.

A last test would be to examine if the velocity component curls are inversely proportional to some `critical Reynolds number' multiplied by the fluid kinematic coefficient, which, as we have seen, might be a quantity proportional to ${\cal D}_t$, in the somewhat turbulent case.

Therefore we propose that our solution, which we currently present as a possible model for chaos (and maybe turbulence) since both its theoretical foundations and its physical properties seem well adapted to such a fluid state, should be put to the tests above described before deciding if it can be accepted or rejected as such a representation. Moreover, these tests should provide a more thorough understanding of the properties exhibited by the model: e. g., the first test, by showing how the transition might occur around some `critical Reynolds number'.

Now, we want to stress that our results have been obtained from the simplest geodesic equation representative of inertial motion in the `fractal space-time' with no additional force or potential nor vector or other (e. g., tensor) field. We can therefore expect that, if this proposal was validated as a representation of turbulence, it should be designed to model some turbulence regime of the simplest kind. Now, depending on the equation of state of the fluid, the `quantum potential' can be regarded as a pressure potential or a mechanical stress tensor. Therefore, we suspect that more complex behaviors should emerge from a motion equation of the kind we have derived and to which new terms would be added that might be obtained from richer properties of the scale relativity construction (an example of such an enrichment is the passage from a simple Euler equation with a `quantum' potential issued from a linear Schr\"odinger equation to a Navier-Stokes equation with pressure and viscosity terms corresponding to a non-linear Schr\"odinger equation with a vector field \cite{LN09}). This will be the subject of future work. However, the three fluid  representation should still obtain in such enlarged cases since it derives from the quaternionic nature of the equations which is itself a product of the symmetry breakings.

Note that our new representation is a generalization of previous proposals aimed at modeling vorticity and/or turbulence.

Fluid vorticity first emerged in our context from a Madelung transformation of a Schr\"odinger equation where the complex wave function was subjected to a vectorial (electromagnetic-type) field \cite{TT52,LN09}. However, the equations which were obtained involved only one rotational velocity field.

A phenomenological two-fluid model of the truncated Euler equations was proposed in \cite{KB07} as a turbulence model. Its results were shown to fit well the predictions of the Eddy-Damped Quasi-Normal Markovian (EDQNM) theory \cite{SO73} which is known to well reproduce the dynamics of the truncated Euler equations \cite{BB06}. Actually, the solutions of these truncated Euler equations have been shown to obey approximately K41 scaling \cite{CB05}. However, the K41 scaling does not reproduce large amplitude events typical of intermittency which is one among the still misunderstood properties of turbulence \cite{MR97}.

Another well-known multi-fluid representation is the two fluid model for Helium II at non-zero temperature. It was developed by Landau \cite{LL41} and involves a `superfluid flow' and a `normal flow' . However, it must be stressed that in such a model the fluid cannot be considered as an actual `mixture' of two different liquids, in a classical sense, since no friction nor momentum transfer occurs between the two parts. One can speak, in a certain sense, of the `superfluid' and of the `normal' parts of the fluid, but this does not mean that it can be actually separated into two such parts \cite{LL59}. We suspect that this kind of feature might apply to our proposal where the `true' nature of the three `flows' might be a same kind of `entanglement'. However, this also remains to be tested.

Now we hope that, if it were validated by experiment, a three fluid representation emerging from the most basic and natural principles and equations of physics might somehow improve our knowledge of complex fluid motions. However, we are aware that solutions of the equations provided here might be hard to find, probably harder than for standard Euler and Navier-Stokes equations which are already actual puzzles for mathematicians. Our hope is however that physical understanding might also gain from the study of the simplest special cases which should be easier to handle.

{\it Aknowledgments.} The author wants to thank Thierry Lehner for valuable discussions and comments about this work and Laurent Nottale for the communication of material and results unpublished at the time this article was submitted for publication.




\begin{thebibliography}{99}

\bibitem{EM27} Madelung, E. (1927). \textit{Zeit. F. Phys.} \textbf{40}, 322.

\bibitem{DB52} Bohm, D. (1952) \textit{Phys. Rev.} \textbf{85}, 166.

\bibitem{TT52} Takabayasi, T. (1952) \textit{Progr. Theor. Phys.} \textbf{8}, 143.

\bibitem{TT53} Takabayasi, T. (1953) \textit{Progr. Theor. Phys.} \textbf{9}, 187.

\bibitem{LN93}  Nottale, L. (1993) \textit{Fractal Space-Time and Microphysics: Towards a Theory of Scale Relativity}, World Scientific, Singapore.

\bibitem{CN04} C\'el\'erier, M.-N. and Nottale, L. (2004) \textit{J. Phys. A: Math. Gen.} \textbf{37}, 931.

\bibitem{LN96b} Nottale, L. (1996) \textit{Chaos, Solitons \& Fractals} \textbf{7}, 877.

\bibitem{CN06} C\'el\'erier, M.-N. and Nottale, L. (2006) \textit{J. Phys. A: Math. Gen.} \textbf{39}, 12565.

\bibitem{LN96a} Nottale, L. (1996) \textit{Astron. Astrophys. Lett.} \textbf{315}, L9.

\bibitem{NS97} Nottale, L, Schumacher G and Gay J 1997 \textit{Astron. Astrophys.} \textbf{322} 1018

\bibitem{NS98} Nottale, L. and Schumacher, G. (1998) in \textit{Fractals and beyond: complexities in the sciences}, Ed. Novak, M. M., World Scientific, p. 149.

\bibitem{DN03} Da Rocha, D. and Nottale, L. (2003) \textit{Chaos, Solitons \& Fractals} \textbf{16}, 565.

\bibitem{CN03}  C\'el\'erier, M.-N. and Nottale, L. (2003) \textit{Electromagnetic Phenomena} \textbf{3}, 83.

\bibitem{NC07} Nottale, L. and C\'el\'erier, M.-N. (2007) \textit{J. Phys. A: Math. Theor.} \textbf{40}, 14471.

\bibitem{LN09} Nottale, L. (2009) \textit{J. Phys. A: Math. Theor.} \textbf{42}, 275306.

\bibitem{CL29} Lanczos, C. (1929) \textit{Z. Physik} \textbf{57}, 447.

\bibitem{AC37} Conway, A. W. (1937) \textit{Proc. Roy. Soc.} {\bf A 162}, 145.

\bibitem{LB04} Love, P. J. and Boghosian, B. M. (2004) \textit{Physica} \textbf{A 332}, 47.

\bibitem{LN05} Nottale, L. (2005) \textit{Progr. Phys.} \textbf{1}, 12.

\bibitem{LN94b} Nottale, L. (1994) invited conference in \textit{Relativity in General} (1993 Spanish Relativity Meeting, Salas), Eds. Diaz Alonso, J. and Lorento Paramo, M., Fronti\`eres, p. 121.

\bibitem{LN99} Nottale, L. (1999) \textit{Chaos, Solitons \& Fractals} \textbf{10}, 459.

\bibitem{MR97} Rieutord, M. (1997) \textit{Une introduction \`a la dynamique des fluides}, Masson, Paris.

\bibitem{AK99} Arnold, V. I. and Khesin, B. A. (1999) \textit{Topological Methods in Hydrodynamics}, Springer-Verlag, New-York.

\bibitem{DD04} Da Rocha, D. (2004) PhD Thesis.

\bibitem{KB07} Krstulovic, G. and Brachet, M.-E. (2007) arXiv: 0710.2462 [physics.flu-dyn]

\bibitem{SO73} Orszag, S. A. (1977) in \textit{Fluid Dynamics} (Proceedings of the Les Houches Summer School 1973), Eds. Balian, R. and Peube, J.-L., Gordon and Breach, New-York, p.235.

\bibitem{BB06} Bos, W. J. T. and Bertoglio, J.-P. (2006) \textit{Phys. Fluids} \textbf{18}, 071701.

\bibitem{CB05} Cichowlas, C., Bona\"iti, P., Debbasch, F. and Brachet, M. (2005) \textit{Phys. Rev. Lett.} \textbf{95}, 264502.

\bibitem{LL41} Landau, L. (1941) \textit{ J. Phys. USSR} \textbf{5}, 71.

\bibitem{LL59} Landau, L. D. and Lifshitz, E. M. (1959) \textit{Fluid Mechanics}, Pergamon Press, Oxford.

\end{thebibliography}
\end{document}